\documentstyle[12pt,aasms4,epsf]{article}

\newcommand{\apgt}{\ {\raise-.5ex\hbox{$\buildrel>\over\sim$}}\ }
\newcommand{\aplt}{\ {\raise-.5ex\hbox{$\buildrel<\over\sim$}}\ }
\newcommand{\eff}{ef\!f}

\begin{document}
\title{Accretion Disks: Limit Cycles and Instabilities}
\author{Mario Livio}
\affil{Space Telescope Science Institute\\
3700 San Martin Drive, Baltimore, MD 21218\\
email: mlivio@stsci.edu}
\begin{abstract}
The existing disk instability model for dwarf nova eruptions is reviewed, in 
the light of recent progress in the understanding of angular momentum 
transport in accretion disks. It is proposed that the standard lower
branch in the ``S-curve'' in the effective temperature-surface density
plane may not exist. Rather, angular momentum transport may be suppressed
in quiescence as a result of cooling. The model for superoutbursts is
also examined, and it is pointed out that recent simulations strongly
support the idea of a thermal-tidal instability.
\end{abstract}
\newpage
\section{Introduction}
The main goal of any discussion of limit cycles in accretion disks is to
explain behaviors like dwarf nova eruptions, the outbursts of black-hole 
x-ray transients, and possibly the outbursts of FU~Orionis stars. The first 
of these is best manifested in the century-long light curve of SS~Cyg (e.g.\ 
Cannizzo \& Mattei 1992; Fig.~1), while the second is extensively reviewed by 
Chen, Shrader \& Livio (1997). At the next level, additional phenomena, like 
the ``standstills'' of Z~Cam systems (e.g.\ AAVSO observations, Fig.~2), and 
the superoutbursts of SU~UMa systems (Fig.~3) must be considered.
\begin{figure}
\centerline{\epsfxsize=4in\epsfbox{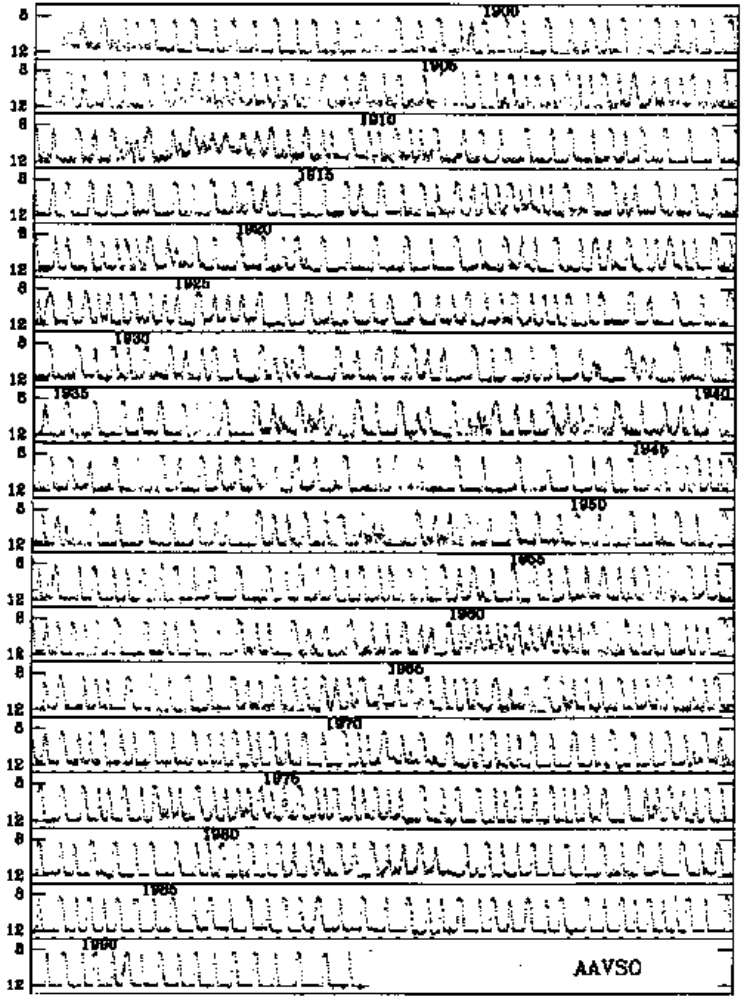}}
\caption{The light curve of SS~Cyg on the basis of AAVSO observations 
(from Cannizzo \& Mattei 1992).}
\end{figure}
\begin{figure}
\centerline{\epsfxsize=4in\epsfbox{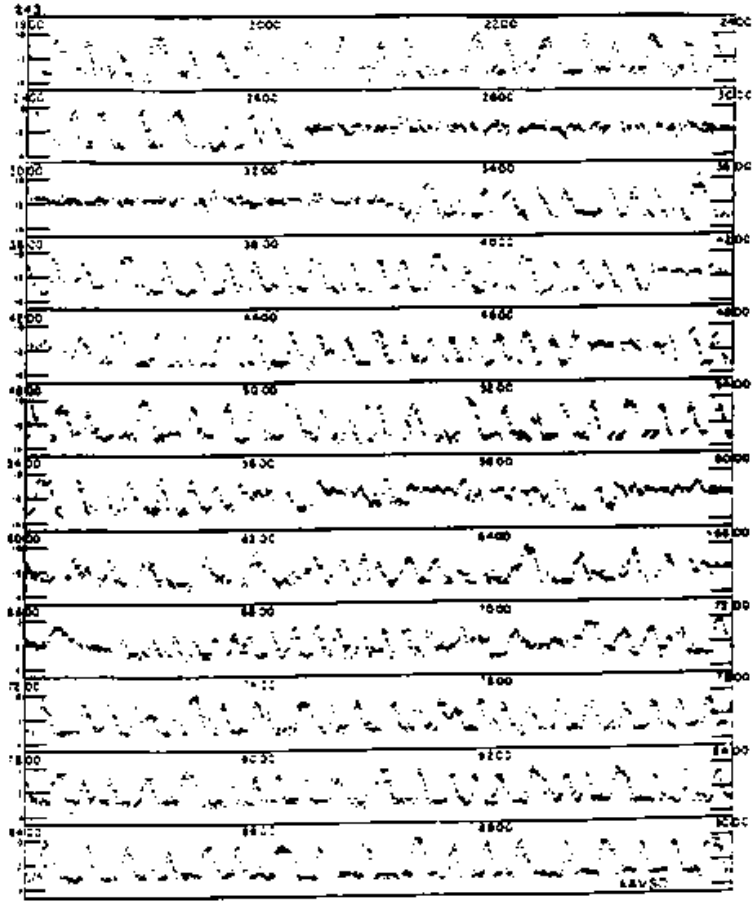}}
\caption{The light curve of Z~Cam on the basis of AAVSO observations
(from Warner 1995).}
\end{figure}
\begin{figure}
\centerline{\epsfxsize=4in\epsfbox{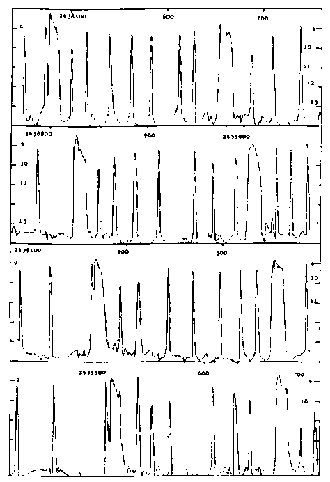}}
\caption{The light curve of VW~Hyi, exhibiting normal eruptions and 
superoutbursts (from Warner 1995).}
\end{figure}

It is generally believed that all of the above phenomena are related to
disk instabilities, with the principal behavior (dwarf nova and x-ray
transient eruptions) being explained in terms of a thermal-viscous disk
instability model (see e.g.\ reviews by Cannizzo 1993; Osaki 1996;
Lasota \& Hameury 1998).

In the present work, I briefly examine some aspects of this and related 
models and I propose a few possible modifications.

\section{The Disk Instability Model}
\subsection{Steady State}

In most of the studies computing limit cycles in standard, thin accretion disks,
the conservation equations are written by averaging over the disk
thickness. For dwarf nova systems these equations take the form (e.g.\
Pringle 1981):\\
mass conservation
\begin{equation}
2\pi R \Sigma (-V_R) = \dot{M}
\end{equation}
conservation of angular momentum
\begin{equation}
\nu\Sigma{d\Omega\over dR} =
\Sigma V_R\Omega +
{{\dot{M}(GM_*R_*)^{1\over2}}\over{2\pi R^3}}
\end{equation}
conservation of energy
\begin{equation}
{{3GM_*\dot{M}}\over{8\pi R^3}}
\left[ 1 - 
\left( {R_*\over R}\right)^{1\over2} \right] = 
\sigma T_{\eff}^4~~.
\end{equation}
Here $\Sigma$ is the disk surface density, $V_R$ is the radial velocity,
$\dot{M}$ is the accretion rate, $\Omega$ is the angular velocity, $\nu$
is the kinematic viscosity and $M_*, R_*$ are the mass and radius
(respectively) of the central object. Combining equations~(1) and~(2)
for a Keplerian flow gives
\begin{equation}
\nu\Sigma =
{\dot{M}\over3\pi}
\left[1 - \left( {R_*\over R}\right)^{1\over2} \right]~~.
\end{equation}

Thus, approximately we have $\dot{M}\sim\nu\Sigma\sim T_{\eff}^4$, a
relation that will become useful later. Equations~(1), (3), and~(4) are
normally augmented by a description of viscosity, which in the
dimensional-analysis
``$\alpha$~prescription'' (Shakura \& Sunyaev 1973) is given by
$\nu=\alpha c_S H$, where $c_S$ is the speed of sound, $H$ is the
vertical disk half-thickness and $\alpha$ is a dimensionless parameter
assumed to satisfy $\alpha\aplt1$ (see \S2.5 and \S3).

\subsection{The Basic Local Limit Cycle}

First, it is important to note that there are at least three fundamental
timescales that are associated with standard accretion disks; dynamical, thermal
and viscous. (A fourth timescale is associated with the propagation of 
transition fronts, see \S 2.4.)  The dynamical time is the period of a 
Keplerian revolution. From hydrostatic equilibrium in the vertical direction, 
it is also the sound crossing time of the disk thickness (or the response 
time to a perturbation of vertical hydrostatics). The thermal timescale 
is the ratio of the thermal content to the local dissipation rate, and the
viscous timescale is the time it takes material to viscously drift
inwards. These timescales are given by:
\begin{eqnarray}
t_{dyn} & \sim & \Omega^{-1}\sim{R\over V_{\phi}}\sim{H\over
c_S}\nonumber\\
t_{th} & \sim & {\Sigma c_S^2\over D(R)}\sim {\Sigma
c_S^2\over\nu\Sigma\Omega^2}\sim{1\over\alpha} t_{dyn}\\
t_{vis} & \sim & {R^2\over\nu}\sim{R^2\over\alpha c_S
H}\sim{1\over\alpha} \left({R\over H}\right)^2 t_{dyn}~~.\nonumber
\end{eqnarray}

Here $D(R)$ is the rate of viscous dissipation (per unit area) and all
other symbols have their usual meaning. As we can see, since
$\alpha\aplt1$ and in standard thin disks $H/R\ll 1$, in such disks
$t_{dyn}<t_{th}<t_{vis}$.

\begin{figure}
\centerline{\epsfxsize=4in\epsfbox{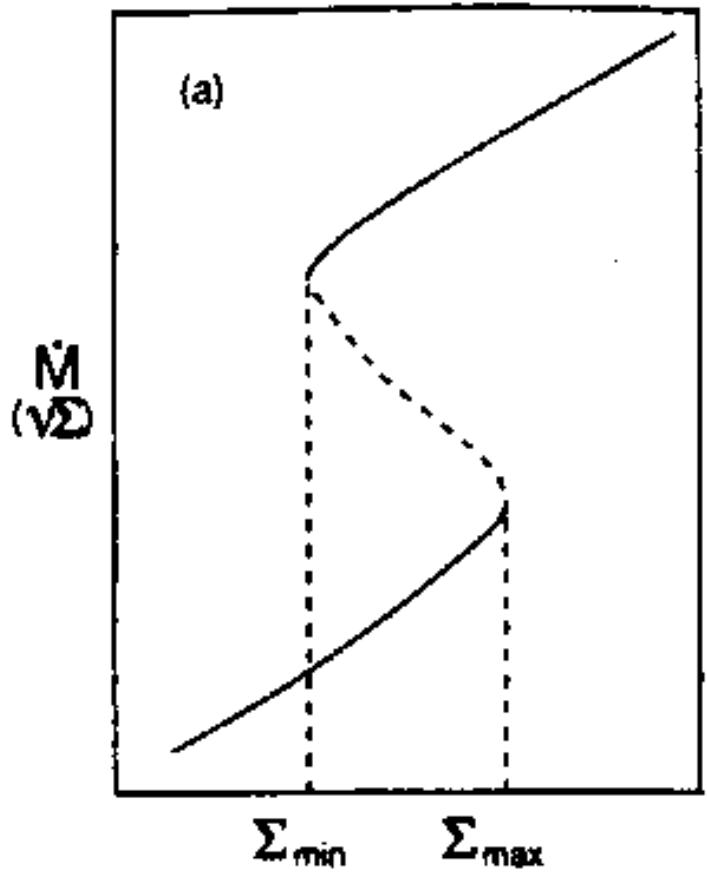}}
\caption{The S-shaped thermal equilibrium curve in the surface
density-effective temperature plane (adapted from Osaki 1996).}
\end{figure}

The steady state equations~(1), (3), and (4) are normally complemented by
equations of hydrostatic equilibrium, mass conservation, and radiative
transfer in the vertical direction. The computations of the local
vertical structure (e.g.\ Meyer \& Meyer-Hofmeister 1981; Cannizzo,
Ghosh \& Wheeler 1982; Smak 1982; Faulkner, Lin \& Papaloizou 1983) were
found to produce a multi-valued thermal equilibrium function in the
(Log~$T_{\eff}$; Log~$\Sigma$) plane (see Fig.~4; from the
discussion following eq.~4, $T_{\eff}$ can also be replaced by $\dot{M}$ or
$\nu\Sigma$). Note that this represents a local solution, at a given radial
distance in the disk. The upper branch of the ``S~curve'' corresponds to
a hot, ionized state of the gas, while the lower branch corresponds to a
cool, neutral state. It is easy to see that the lower and upper branches
are stable (e.g.\ an increase in the surface density, which results in
an increase in the viscous energy production rate, leads to an increase
in the effective temperature, and concomitantly in the energy loss
rate), while the middle branch is not. Imagine now that the rate at
which mass is being transferred from the companion star is such that it
corresponds to a point on the middle, unstable branch for some annulus
(see Fig.~4). In such a case, no local stable equilibrium is possible.
Rather, because the rate at which mass is being supplied is higher than the
rate at which it can be transported locally, the surface density will
increase (along the lower curve) until the critical surface density
$\Sigma_{max}$ is reached, at which point the annulus will heat up on a
thermal timescale, ``jumping'' to the upper branch. There, since mass is
being transported faster than it is being supplied, the surface density
will decrease (along the upper curve) until the critical density
$\Sigma_{min}$ is reached. At that point the annulus will jump back to
the lower branch, thus completing the limit cycle.

Before turning to time dependent evolution it is instructive to examine
the physical reasons for the existence of such a multi-valued S-curve.

\subsection{The S-Curve}

The physical reason for the S-shaped equilibrium curve in the
($\nu\Sigma, \Sigma$) (or ($T_{\eff},\Sigma$)) plane is the partial ionization 
of hydrogen at $T\sim10^4$~K (e.g.\ Mineshige \& Osaki 1983; Pojmanski 1986; 
Cannizzo 1992; Liu \& Meyer-Hofmeister 1997).

\begin{figure}
\centerline{\epsfxsize=4in\epsfbox{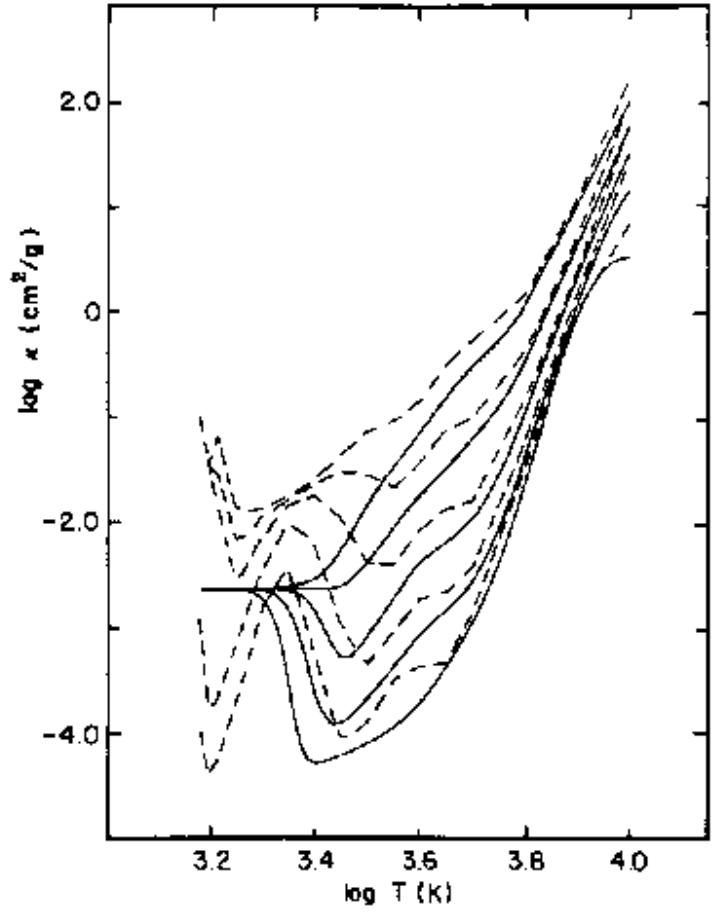}}
\caption{Rosseland mean opacities compiled by Cannizzo \& Wheeler (1984).}
\end{figure}
\begin{figure}
\centerline{\epsfxsize=3in\epsfbox{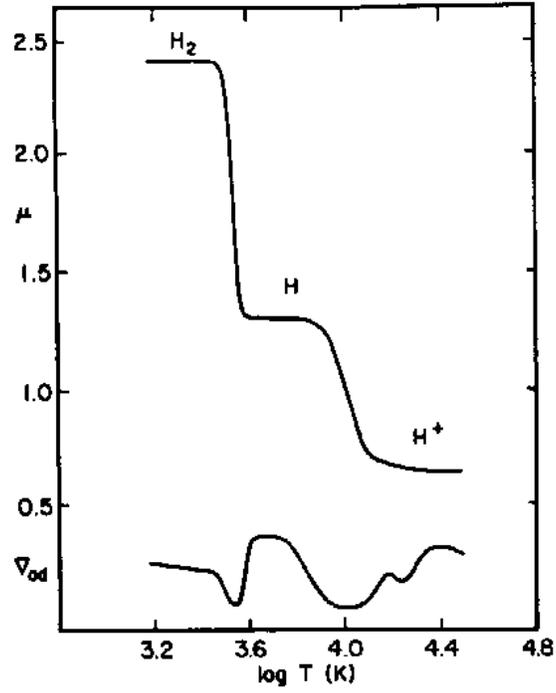}}
\caption{The mean molecular weight $\mu$ and the adiabatic index 
$\bigtriangledown_{ad}$. Note the decrease in $\bigtriangledown_{ad}$ due to 
partial ionization (from Cannizzo \& Wheeler 1984).}
\end{figure}
\begin{figure}
\centerline{\epsfxsize=3in\epsfbox{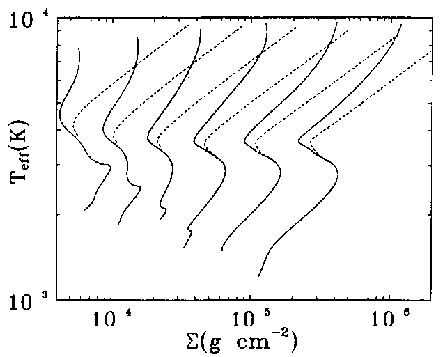}}
\caption{The effective temperature vs.\ surface density for (from left to 
right) $\log\alpha=0.5$, 0, --0.5, --1.0, --1.5 and --2.0. From Cannizzo 
(1992).}
\end{figure}

This partial ionization produces two effects which give the S-curve its
shape: (i)~there is a steep dependence of the opacity on the temperature
in the range $T\sim6000$--10000~K (Fig.~5), which causes a change in the
sign of $d\log (\nu\Sigma)/d\log \Sigma$, and (ii)~the adiabatic
temperature gradient drops substantially (Fig.~6), thus driving
convection. Both of these effects produce a local maximum in log~$\Sigma$ as 
a function of log~$T_{\eff}$ (Pojmanski 1986; Cannizzo 1992; See
Fig.~7). Here, however, arises a difficulty which in my opinion has not
been fully appreciated by disk instability modellers.  While the opacity
peak is hardly significant (e.g.\ in Fig.~7 it appears only for rather
high ($\alpha\sim1$), possibly unrealistic values of the viscosity
parameter), it is the convection peak which determines the location, and
to some extent the existence of the lower branch. This fact {\it raises
serious doubts about the reality of the lower branch\/}. It should be
realized that it appears almost certain now that angular momentum transport 
and energy dissipation in disks are governed by MHD turbulence (e.g.\ Hawley, 
Gammie \& Balbus 1996; Brandenburg, et~al.\ 1995). Under these circumstances,
and with the possibility that much of the energy dissipation occurs in a disk 
corona, the role of standard (mixing length) convection is questionable at 
best. I will return to this problem in \S3.

\subsection{Time Dependent Evolution}

Many time-dependent simulations of the disk evolution have been carried out
by a number of researchers; the most recent ones being by Cannizzo (1998), 
Hameury et~al.\ (1998) and Menou, Hameury \& Stehle (1998).

For the time-dependent case, the conservation equations can be
written as:\\
mass conservation
\begin{equation}
{\partial\Sigma\over\partial t} =
-{1\over r} {\partial\over\partial r} (r\Sigma V_r) +
{1\over 2\pi r} {\partial\dot{M}_{tr}\over 2r}
\end{equation}
angular momentum conservation
\begin{equation}
{\partial(\Sigma r^2\Omega)\over\partial t} =
-{1\over r} {\partial\over\partial r} (r^3 \Omega\Sigma V_r) +
{1\over r} {\partial\over\partial r}
\left( r^3 \nu\Sigma {d\Omega\over d\Sigma}\right) +
{j_s\over 2\pi r} {\partial\dot{M}_{tr}\over\partial r} -
{1\over 2\pi r} \tau_{tid}(r)
\end{equation}
energy conservation
\begin{equation}
c_p\Sigma {\partial T\over\partial t} =
2(Q^+ - Q^-) - 2{H\over r} {\partial\over\partial r} (r F_r) -
c_p V_r\Sigma {\partial T\over\partial r} -
{{\cal R} T\over\mu} {\Sigma\over r} {\partial (r V_r)\over \partial r}
\end{equation}

Here $\dot{M}_{tr}$ is the rate at which mass is incorporated into the
disk, $j_s$ is the specific angular momentum of the stream from the
inner Lagrangian point, $\tau_{tid}$ is the tidal torque applied by the
secondary star, $Q^+$ and $Q^-$ are the rates of heating and cooling
(per unit area) respectively, and all other symbols have their
usual meaning. While some small differences in the exact expressions
used for the various terms exist among different researchers, these do
not appear to be crucial for the general behavior.

The broad-brush evolution can be described as follows. In quiescence
matter piles up in the disk, until at some radius the surface density
exceeds $\Sigma_{max}$ (Fig.~4). The corresponding annulus heats up on a
thermal timescale. Since the viscosity increases with temperature (even
for a constant $\alpha$), this annulus starts to spread, thus initiating
the propagation of heating fronts, which eventually bring the entire
disk to a high state (the upper branch). A typical evolution of the
surface density, the accretion rate onto the central object and the disk
mass is shown in Fig.~8 (taken from Mineshinge \& Osaki 1985 and
Cannizzo, Wheeler \& Polidan 1986). Two cases are shown; one in which
the outburst begins at a large radius, and one in which it begins at a
small radius. As can be seen in the figures, since most of the disk mass
lies in the outer parts of the disk, outbursts starting from the outside
produce a much steeper increase in the accretion rate. Since in the high
state $\Sigma(r)\sim r^{-1}$, while $\Sigma_{min}\sim r$, it is always
the case that $\Sigma$ first drops below $\Sigma_{min}$ at the outer
disk radius. This initiates the propagation of a cooling front which
transforms the entire disk to a low state configuration, thus
terminating the outburst. In cases in which the cooling front cannot
propagate (e.g.\ due to strong irradiation of the disk by the central
source), the evolution is slower, because it has to proceed on the
viscous timescale. This may produce flat-topped maxima in the light curve.
\begin{figure}
\centerline{\epsfxsize=3in\epsfbox{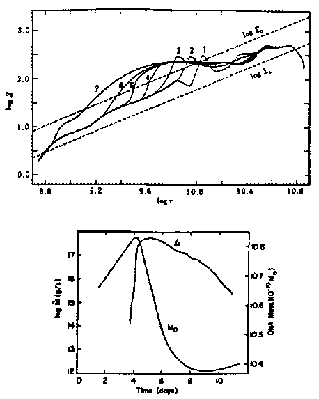}\epsfxsize=3.1in\epsfbox{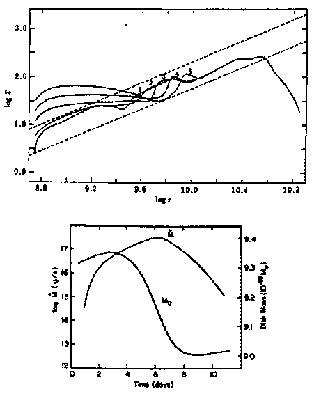}}
\caption{The left two figures (8a) show how the disk responds to an outburst 
which begins at a large radius. The right two figures (8b) show an outburst 
starting at a small radius. The top figures show the evolution of the surface
density, the bottom two figures show the accretion rate and the disk mass
(adapted from Cannizzo 1993).}
\end{figure}

The structure and physical properties of the transition fronts have been
studied extensively both analytically and numerically (e.g.\ Meyer 1984;
Papaloizou \& Pringle 1985; Lin, Papaloizou \& Faulkner 1985; Cannizzo,
Chen \& Livio 1995; Vishniac \& Wheeler 1996; Vishniac 1997; Menou,
Hameury \& Stehle 1998). Broadly speaking, these works have shown that
the speed of the fronts is of the order of $V_F\sim\alpha_{H} c_S$,
where $\alpha_{H}$ is the viscosity parameter in the hot state (see
below) and $c_S$ is the speed of sound in the front (for a more detailed
discussion see Vishniac 1997). The width of the front was found by Menou 
et~al.\ (1998) to be proportional to the disk pressure scale height $H$.

\subsection{Viscosity}

Angular momentum transport in accretion disks is due to magnetically
driven turbulence (e.g.\ Balbus \& Hawley 1998; Godon \& Livio 1998 and 
references therein).
Consequently, expressing the anomalous viscosity by means of a fixed
parameter does not represent adequately the physical situation. The
following discussion should be viewed therefore as a time and space
overaging of the process of angular momentum transport. Since the MHD
simulations indicate that Maxwell stresses dominate over Reynolds
stresses (e.g.\ Hawley et~al.\ 1996), the viscosity parameter $\alpha$
is given by the appropriately overaged value of $B_R B_{\phi}/4\pi
\rho c_S^2$, where $B_R$ and $B_{\phi}$ are the radial and toroidal
(respectively) components of the dynamo-generated magnetic field.

From disk instability models of dwarf nova eruptions, it was generally
found that in order to reproduce the observed amplitudes and timescales
one needs to assume that the viscosity parameter in the hot state,
$\alpha_{H}$, is larger than the one in the cold state,
$\alpha_{C}$, by a factor of 5--10 (e.g.\ Smak 1984). Otherwise, only
relatively frequent, small oscillations in the luminosity are produced.
The changes to a local S-curve, introduced by using
$\alpha_{H}/\alpha_{C}\sim5$ instead of a constant value, are shown
in Fig.~9 (Hameury et~al.\ 1998).
\begin{figure}
\centerline{\epsfxsize=4in\epsfbox{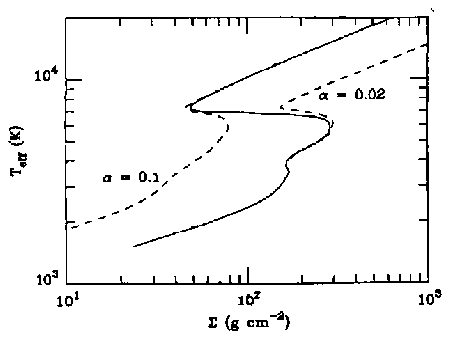}}
\caption{The dashed S-curves correspond to constant $\alpha$, 
$\alpha=\alpha_H$ (left) and $\alpha=\alpha_C$ (right).  The solid line
represents $\alpha=\alpha_H$ on the upper branch and $\alpha=\alpha_C$ 
on the lower branch (from Hameury et~al.\ 1998).}
\end{figure}

One may ask, do we have any observational handle on the value of $\alpha$?

From visual amplitudes and rates of decline (when compared to models) one 
obtains $\alpha_{H}\sim0.2$ (Smak 1984). On the other hand, Livio \& Spruit 
(1991) have shown that from the recurrence times of dwarf nova eruptions 
one obtains $\alpha_{C}\aplt0.05$ (see also Smak 1996).

\subsection{A Representative Model Calculation}

The most recent, extensive numerical models have been carried out by
Hameury et~al.\ (1998). Unlike in some of the previous calculations,
these authors use an adaptive grid technique and an implicit numerical
scheme, which allows them to resolve the transition fronts. Hameury
et~al.\ also allow the outer disk radius to vary---a boundary condition
which proves to be important for the obtained behavior (see below).  The
results of two typical calculations are shown in Figs.~10--11, for two
values of the accretion rate.  The main difference between the two
calculations is the fact that for the higher mass transfer rate
($\dot{M}=10^{17}$~gs$^{-1}$) the eruption is ``outside-in'' (namely,
starts at a large radius with the transition front propagating inwards;
these are also known as ``type~A'' outbursts, Smak 1987), while for the
lower ($\dot{M}=10^{16}$~gs$^{-1}$) it is ``inside-out'' (type~B).  This
basic difference in the behavior is well understood.  In the case of a high 
mass transfer rate, material accumulates in the outer disk faster than it 
drifts inwards, hence the surface density first exceeds the local
$\Sigma_{max}$ in the outer part. For relatively low accretion rates
$\Sigma_{max}$ is first exceeded in the inner disk. The obtained
behavior of the disk radius agrees well with observations (e.g.\ Smak
1984; O'Donoghue 1986). As the outburst is initiated and material starts
to be accreted, it transports its angular momentum outwards, thus the
outer radius expands (Livio \& Verbunt 1988).
\begin{figure}
\centerline{\epsfxsize=3in\epsfbox{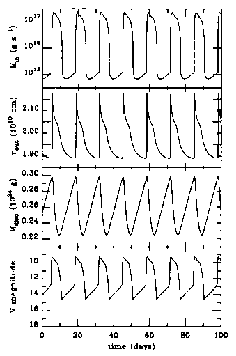}}
\caption{The outbursts obtained for $M_{WD}=0.6$~M$_\odot$, $\alpha_C=0.04$,
$\alpha_H=0.2$ and $\dot{M}=10^{17}$ gs$^{-1}$ (from Hameury et~al.\ 1998).}
\end{figure}
\begin{figure}
\centerline{\epsfxsize=2.75in\epsfbox{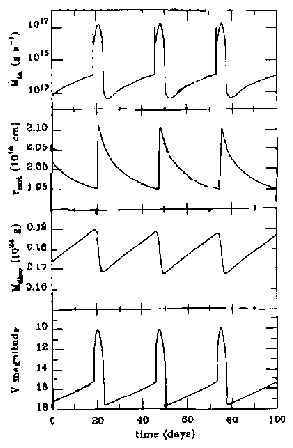}}
\caption{The same as Fig.~10, for $\dot{M}=10^{16}$ gs$^{-1}$.}
\end{figure}

As I noted above, Hameury et~al.\ examined the effects of using as an
outer boundary condition a disk radius, $R_{out}$, that is allowed to
vary (see also Ichikawa \& Osaki 1992; Livio \& Verbunt 1988), as
opposed to a fixed $R_{out}$. They found that when $R_{out}$ was kept
fixed, the outbursts always tended to be inside-out, and they involved
the accretion of a larger fraction of the disk mass. The reason for the
latter behavior is easy to understand. When $R_{out}$ is fixed, the material
that moves outward (with the angular momentum it absorbed) piles up at
the outer edge increasing there the surface density. Consequently, all
of that material needs to be accreted before the surface density drops
below $\Sigma_{min}$ and the cooling wave is initiated. The light curves 
obtained under the two different boundary conditions are shown in Fig.~12.
\begin{figure}
\centerline{\epsfxsize=3.75in\epsfbox{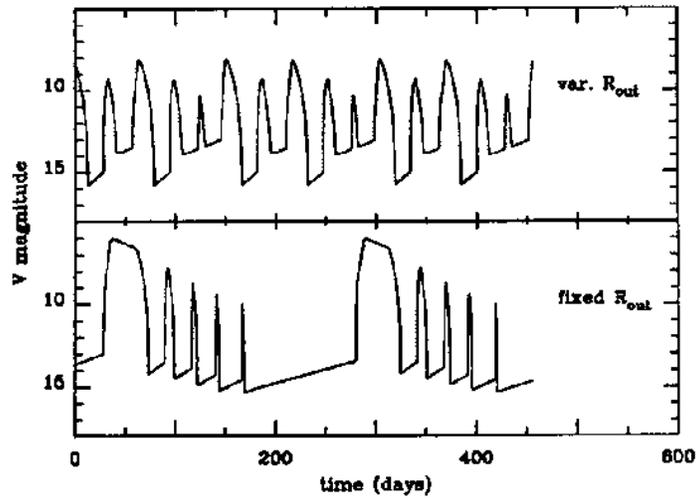}}
\caption{The light curves that are obtained assuming a fixed outer disk
radius (lower panel), and allowing the outer radius to vary (upper panel).
From Hameury et~al.\ (1998).}
\end{figure}

\clearpage
I should note that in addition to successfully reproducing light curves
and the behavior of the disk radius, the disk instability model has 
provided a reasonable explanation for the
observed distribution of systems in stability/instability regions in the 
($\dot{M}$, $P_{orb}$) plane (where $P_{orb}$ is the binary orbital period; 
e.g.\ Smak 1989; Osaki 1996). In particular, the Z~Cam objects which
experience ``standstills'' (Fig.~2) are found to be right at the border
between the unstable and stable regions. Thus, it is expected that small
changes in $\dot{M}$ can carry them into and out of the instability
region, which can account for the standstills.

\section{MHD Turbulence and Limit Cycles}

Since angular momentum transport in accretion disks is driven by MHD
turbulence, it is appropriate to ask whether the thermal limit cycle
model as described in the previous sections is consistent with such a
dynamo-generated ``viscosity.'' In particular, it is important to
attempt to understand what could cause the existence of two values for
the viscosity parameter, namely $\alpha_H$ and $\alpha_C$. In this
respect, the following {\it suggestion\/} has been made (Armitage, Livio
\& Pringle 1996; Gammie \& Menou 1998): {\it the MHD turbulence is
quenched in quiescence, leading to a low viscosity, while it is
operative in outburst, leading to a high viscosity\/}.

The exact mechanism by which the MHD turbulence is suppressed is not
clear at the moment, but there exist at least two possibilities:
(i)~when the disk cools, the magnetic field does not decay as fast (it
decays on a superthermal timescale, e.g.\ Tout \& Pringle 1992).
Consequently, the condition for magnetorotational instability (Balbus \&
Hawley 1991), $B_z^2<24/\pi \rho c_S^2$, is violated, and hence the
dynamo action is suppressed. (ii)~When the disk cools, the magnetic
Reynolds number $Re_M=c_S H/\eta$ (where $\eta$ is the resistivity) can
decrease below $\sim10^4$. Numerical simulations indicate that the MHD
turbulence may be suppressed when $Re_M\aplt10^4$ (Hawley, Gammie \& Balbus
1996).

While I am slightly more inclined to favor the second possibility, the
important thing for the present discussion is the fact that viscosity is
decreased in quiescence as a result of cooling.

If viscosity in quiescence is indeed low, this should have some obvious
observational consequences: (i)~little accretion should take place in
quiescence. (ii)~There should be little or no emission from the boundary
layer and little UV emission from the inner disk. (iii)~The temperature
profile in quiescence should be very different from that of a standard
disk (it should basically be flat). (iv)~A UV delay (with respect to
optical) should be observed in the rise to outburst.

There are strong indications that all of these predictions are
consistent with observations at least in some systems. For example,
observations of all of the black hole transients show very low accretion
in quiescence. Similarly, observations of the cataclysmic variables OY~Car 
and Z~Cha indicate that
the accretion rate in the inner disk is lower by about a factor of 100
than that in the outer disk (e.g.\ Wood et~al.\ 1986, 1989). Also, the
temperature profile in OY~Car is extremely flat in quiescence, while it
is consistent with that of a standard disk in Z~Cha in outburst.

It therefore appears that there is some merit to the suggestion that the
viscosity in quiescence is low. However, an important question arises,
if material accumulates more or less in a ring (due to the low
viscosity), {\it how can inside-out (type~B) outbursts be obtained\/}?
Before I attempt to answer this question it is important to examine the
observational evidence that inside-out outbursts do indeed exist. There are by
now 15~dwarf novae with claimed positively identified outburst types
(e.g.\ Smak 1996; Warner 1995, Table 3.6). Outside-in outbursts are
characterized by a {\it rapid rise\/}, the existence of a {\it UV
delay\/}; and they follow a {\it broad loop\/} in a color-color diagram.
All of these properties follow directly from the fact that the outburst
starts from the outer disk, where most of the mass is concentrated (see \S2.6).
Examples for such outbursts can be seen in VW~Hyi (Smak 1987; Fig.~13)
and they are well reproduced in theoretical models. Inside-out outbursts
are characterized by a {\it slower rise\/}, {\it no UV delay\/}, and a
{\it narrow loop\/} in the color-color diagram (Fig.~13). All of these,
in turn, are a consequence of the fact that the outburst starts in the
inner part of the disk, where there is relatively little mass. Examples
to this type of outburst are seen in SS~Cyg and AH~Her, and again they
are adequately reproduced by models. One can therefore conclude with
some certainty that outbursts of the two types do exist.
\begin{figure}
\centerline{\epsfxsize=4in\epsfbox{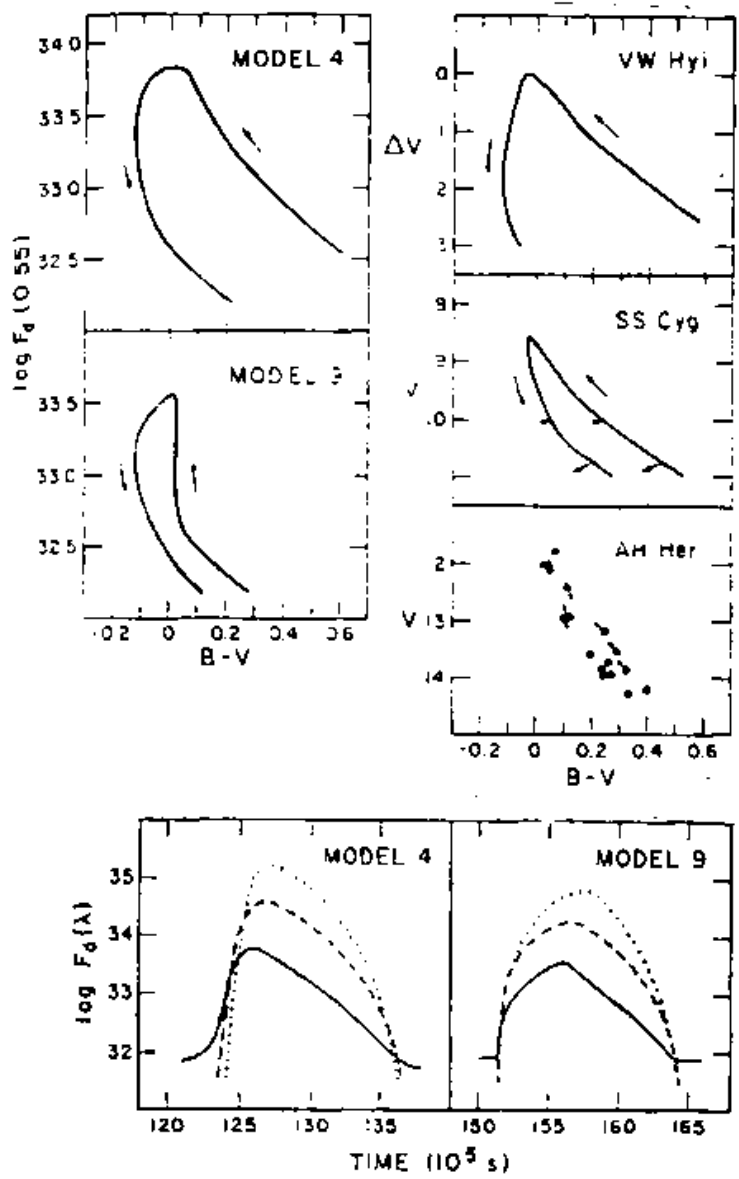}}
\caption{Examples of observations and model calculations showing a rapid
rise and a broad loop in the color-color diagram for outside-in outbursts, 
and a slower rise and a narrow loop in the color-color diagram for inside-out 
outbursts (adopted from Smak 1987).}
\end{figure}

I now return to the question of how can inside-out outbursts be
reconciled with the suggestion of low viscosity at quiescence.  In
particular, I note the following, seemingly strange fact (e.g.\ Warner 1987):
SU~UMa systems show {\it only\/} outside-in outbursts while Z~Cam
systems show {\it only\/} inside-out outbursts, in spite of the fact
that the accretion rate in Z~Cam systems is more than ten times higher
than in SU~UMa systems!

I would like to suggest the following possible solution. If we take as
the critical (for suppression of MHD turbulence) magnetic Reynolds
number $Re_M^{crit}\sim10^4$, then it can be shown (e.g.\ Gammie \&
Menou 1998) that the critical temperature corresponding to this value
(so that for $T>T_{crit}$, $Re_M>Re_M^{crit}$) is approximately given by
\begin{equation}
T_{crit}\sim7400{\rm K}~\Omega^{0.2}\Sigma^{0.038}~~.
\end{equation}
It can easily be shown then that the condition $T>T_{crit}$ assumes the
form $\xi\apgt$const., with
\begin{equation}
\xi=\left({\dot{M}\over P_{orb}}\right)^{0.3} f^{-{3\over4}}(q)~~,
\end{equation}
where $P_{orb}$ is the orbital period and f(q) is some function of
the mass ratio of the two binary components. Now, it turns out that
$\xi_{Z~Cam}\sim2\xi_{SU~U\!Ma}$. Therefore, it is possible that the disk
has a non-negligible viscosity in quiescence in Z~Cam (and similar)
systems and no viscosity in systems like SU~UMa. This could resolve the
puzzle of inside-out outbursts.

I therefore propose as a possibility the following new scenario for dwarf
nova eruptions (see also Armitage, Livio, \& Pringle 1996; Gammie \&
Menou 1998). In the disks of systems in which at quiescence $Re_M\ll
Re_M^{crit}$ (or some equivalent condition for the suppression of MHD
turbulence), there is essentially no viscosity at quiescence. These
systems experience outside-in outbursts. The initiation of these
outbursts may be aided by the ring of accumulated material becoming
unstable to the Papaloizou-Pringle (1994) instability (e.g.\ Rozyczka \&
Spruit 1993).

In the disks of systems in which at quiescence $Re_M\apgt Re_M^{crit}$
(or an equivalent condition allowing MHD turbulence), accretion takes
place also in quiescence. In such systems the possibility for inside-out
eruptions exists. The main implication of this new scenario is that {\it
a ``standard'' lower branch of the S-curve may not exist\/}. This would
be consistent with the doubts I raised about the existence of such a
branch in \S2.3.  It is clear that the viscosity cannot be zero at
quiescence in all systems, since some systems (e.g.\ HT~Cas, WX~Hyi)
exhibit low states which are almost certainly associated with a
reduction in $\dot{M}$ (these low states would not have been seen if
there was no viscosity).

\section{Superhumps and Superoutbursts}

I would like now to discuss briefly another disk instability, the one
producing superoutbursts and superhumps. Superhumps are periodic
photometric humps that are observed in some dwarf nova systems, and they
have periods that are longer by a few percent than the orbital period
(Fig.~14; see e.g.\ Osaki 1996: Warner 1995 for reviews of their
properties). Most of the systems that exhibit superhumps (the SU~UMa
systems) have orbital periods below the 2--3 hr period gap in the
distribution of cataclysmic variables (TU~Men is an exception). Some
black hole soft x-ray transients also apparently exhibit superhumps 
(O'Donoghue \& Charles 1996). In the SU~UMa systems, superhumps are 
observed mostly during superoutbursts (see below), although in some systems 
(like V1159~Ori) they have been observed after superoutbursts (e.g.\ 
Patterson et~al.\ 1995).
\begin{figure}
\centerline{\epsfxsize=3.5in\epsfbox{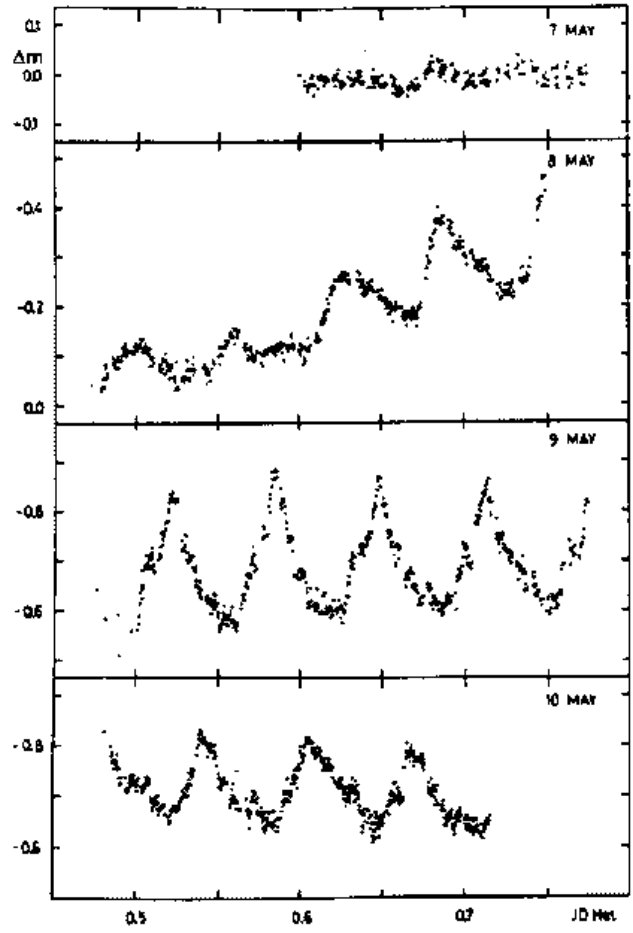}}
\caption{The development of superhumps in V436~Cen. From Semeniuk (1980).}
\end{figure}

Superoutbursts are eruptions that are brighter than normal by about
0.7~mag, and longer than normal by a factor of 5--10 (Fig.~3).  The
general impression is that superoutbursts start with normal outbursts
(e.g.\ Vogt 1974; Warner 1995).

The basic model for superhumps has been suggested by Whitehurst (1988),
and further investigated by Hirose \& Osaki (1990) and Lubow (1991). It
involves a tidally driven eccentric instability of the accretion disk.
The eccentric disk precesses at a period $P_{prec}$. The superhump
period is identified as the synodic period between the precessing
eccentric disk and the orbital period
\begin{equation}
{1\over P_{SH}}={1\over P_{orb}} - {1\over P_{prec}}~~.
\end{equation}

Non-axisymmetric waves in the disk can be expressed in the form $\exp
[i(k\theta - \ell\Omega t)]$ (where $\theta=\phi + \Omega t$, with
$\phi$ being the azimuthal angle in the corotating frame). The modes are
thus identified by the pair ($k, \ell$). Hirose \& Osaki (1990) and in
particular Lubow (1991) demonstrated the crucial role played by the 3:1
resonance in the disk in the excitation of the eccentric instability
(see also Whitehurst \& King 1991; Molnar \& Kobulnicky 1992). Very
briefly, a perturbation in the eccentricity (characterized by
(1,0)) combines with the resonant effect of the 3:1 resonance in the 
tidal potential (corresponding to (3,3)) to produce a two-armed 
spiral density wave (characterized by (2,3)). This spiral wave 
((2,3)), in turn, combines again with the tidal potential 
((3,3)) to amplify the eccentricity ((1,0)). For the 
instability to grow, the 3:1 resonance radius must lie within the disk. This 
implies that this instability can occur only for mass ratios satisfying
$q=m_2/m_1\aplt0.25$. This explains the fact that superhumps are found
primarily below the period gap or in black hole x-ray binaries.

Surprisingly, it has proven much more difficult to find a convincing
model for the superoutbursts.  The history of the field is nicely
summarized in Warner (1995) and Osaki (1996). The most promising model
has been suggested by Osaki (1989). In this model, the following
sequence of events is suggested to occur. As the accretion disk
undergoes normal dwarf nova eruptions, in each outburst the mass that is
actually accreted onto the white dwarf is smaller than the mass
accumulated (via transfer) during the inter-outburst period. This agrees
with the results of detailed simulations (e.g.\ Cannizzo 1993).
Consequently, both the mass and the angular momentum of the disk are
building up gradually, and the disk radius experiences a slow overall
increase with each successive eruption (Fig.~15).
\begin{figure}
\centerline{\epsfxsize=3in\epsfbox{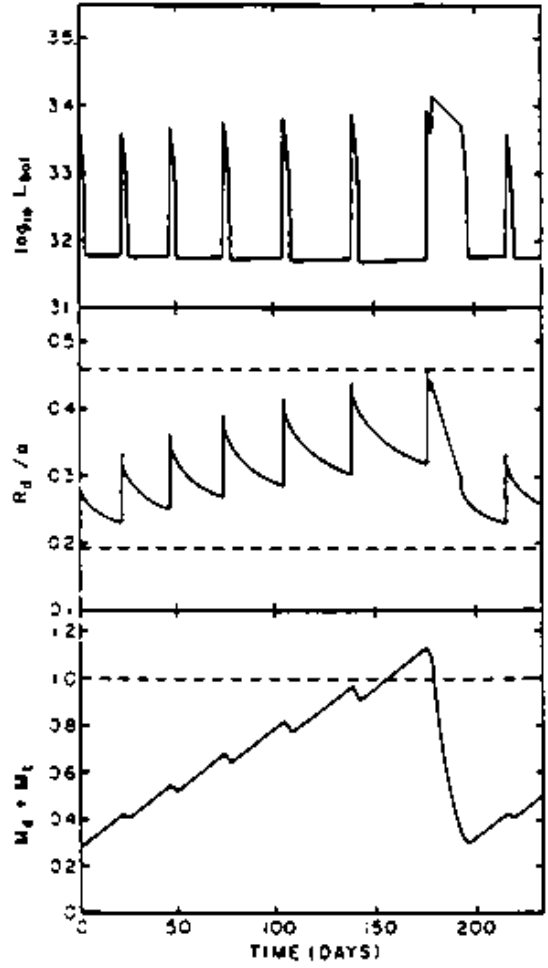}}
\caption{The time evolution of the accretion disk luminosity, radius, and mass
(from top to bottom) in a simplified thermal-tidal instability model for 
superoutbursts (from Osaki 1996).}
\end{figure}

At some point, the disk radius hits the 3:1 resonance and the  tidally
induced eccentric instability sets in, producing a precessing eccentric
disk. Osaki {\it assumed\/} that the eruption that ensues at that point
results in an {\it enhanced\/} rate of removal of angular momentum by
the tidal interaction (compared to normal outbursts), so that a much
larger fraction of the disk mass is accreted, thus producing a
superoutburst. Until recently,  all of the suggestions in Osaki's model
have been verified by numerical simulations, except the assumption about
an increased rate of removal of angular momentum. Most recently however,
a simulation by Murray (1998) showed that this assumption is likely to be
correct too. Murray simulated normal outbursts by an increased
viscosity and showed that once the disk encounters the 3:1 resonance it
becomes tidally unstable and the tidal torques become much more
efficient in removing angular momentum from the disk. He further showed
that the resulting increased accretion rate is consistent with
superoutbursts.

\section{Tentative Conclusions}

In recent years it appeared as if the question of dwarf nova eruptions
has been fully solved, maybe with just a few details left to be sorted
out. On the other hand, on the question of superoutbursts considerable
skepticism existed in relation to Osaki's model. The enormous progress
in the understanding of angular momentum transport in accretion disks on
one hand, and the increased sophistication of numerical simulations on
the other, have contributed to a renewed interest in both of these
problems.

On the basis of the material presented in the previous sections I am
inclined to make the following tentative suggestions:

(1) Dynamo generation of viscosity may be suppressed in some systems
(and reduced in others) in quiescence, due to cooling. The difference in
the angular momentum transport properties may be the main cause for
dwarf nova eruptions and x-ray transient outbursts.

A continuing serious effort will be required, in MHD simulations on one hand, 
and in observations of systems in quiescence on the other, to determine the 
viability of the above suggestion.

(2) Superoutbursts are probably caused by a thermal-tidal instability,
as has been proposed by Osaki (with the normal outbursts however occurring as 
in (1) above). 

The best observational test of this model can be provided by {\it monitoring
the behavior of the disk radius\/} through a series of normal outburst and a
superoutburst.

\begin{acknowledgements}
This work has been supported by NASA Grant NAG5-6857. I thank the Isaac
Newton Institute for Mathematical Sciences for its hospitality and John
Cannizzo and Kristen Menou for helpful discussions.
\end{acknowledgements}


\begin{references}

\reference{}Armitage, P.~J., Livio, M.\ \& Pringle, J.~E. 1996, ApJ,
457, 332

\reference{}Balbus, S.~A.\ \& Hawley, J.~F. 1991, ApJ, 376, 214

\reference{}Balbus, S.~A.\ \& Hawley, J.~F. 1998, Rev.\ Mod.\ Phys., 70,
1

\reference{}Brandenburg, A., Nordland, A., Stein, R.~F.\ \& Torkelsson,
U. 1995, ApJ, 446, 741

\reference{}Cannizzo, J.~K. 1992, ApJ, 385, 94

\reference{}Cannizzo, J.~K. 1993, in Accretion Disks in Compact Stellar
Systems, ed.\ J.~C.\ Wheeler (Singapore: World Scientific), p.~6

\reference{}Cannizzo, J.~K. 1998, ApJ, 494, 366

\reference{}Cannizzo, J.~K., Ghosh, P.\ \& Wheeler, J.~C. 1982, ApJ, 260,
L83

\reference{}Cannizzo, J.~K.\ \& Mattei, J. 1992, ApJ, 401, 642

\reference{}Cannizzo, J.~K.\ \& Wheeler, J.~C. 1984, ApJS, 55, 367

\reference{}Cannizzo, J.~K., Wheeler, J.~C.\ \& Polidan, R.~S. 1986,
ApJ, 301, 634

\reference{}Chen, W., Shrader, C.~R.\ \& Livio, M. 1997, ApJ, 491, 312

\reference{}Faulkner, J., Lin, D.~N.~C.\ \& Papaloizou, J. 1983, MNRAS,
205, 359.

\reference{}Gammie, C.~F. \& Menou, K. 1998, ApJ, 492, L75

\reference{}Godon, P.\ \& Livio, M. 1998, ApJ, submitted

\reference{}Hameury, J.-M., Menou, K., Dubus, G., Lasota, J.-P.\ \&
Hur\'e, J.-M. 1998, MNRAS, in press, Astro-ph/9803242

\reference{}Hawley, J., Gammie, C.~F.\ \& Balbus, S. 1996, ApJ, 464, 690

\reference{}Hirose, M.\ \& Osaki, Y. 1990, PASJ, 42, 135

\reference{}Ichikawa, S.\ \& Osaki, Y. 1992, PASJ, 44, 15

\reference{}Lasota, J.-P.\ \& Hameury, J.-M. 1998, in Accretion
Processes in Astrophysics---Some Like it Hot, eds.\ S.~Holt \&
T.~Kallman, in press, Astro-ph/9712202

\reference{}Lin, D.~N.~C., Papaloizou, J.~C.~B.\ \& Faulkner, J. 1985,
MNRAS, 212, 105

\reference{}Liu, B.~F.\ \& Meyer-Hofmeister, E. 1997, A\&A, 328, 243

\reference{}Livio, M. \& Spruit, H.~C. 1991, A\&A, 252, 189

\reference{}Livio, M.\ \& Verbunt, F. 1988, MNRAS, 232, 1p

\reference{}Lubow, S. 1991, ApJ, 381 259

\reference{}Menou, K., Hameury, J.-M.\ \& Stehle, R. 1998, MNRAS, in
press

\reference{}Meyer, F. 1984, A\&A, 131, 303

\reference{}Meyer, F.\ \& Meyer-Hofmeister, E. 1981, A\&A, 104, L10

\reference{}Mineshige, S.\ \& Osaki, Y. 1983, PASJ, 35, 377

\reference{}Molnar, L.~A.\ \& Kobulnicky, H.~A. 1992, ApJ, 392, 678

\reference{}Murray, J.~R. 1998, MNRAS, 297, 323

\reference{}O'Donoghue, D. 1986, MNRAS, 220, 23p

\reference{}O'Donoghue, D.\ \& Charles, P.~A. 1996, MNRAS, 282, 191

\reference{}Osaki, Y. 1989, PASJ, 41, 1005

\reference{}Osaki, Y. 1996, PASP, 108, 39

\reference{}Papaloizou, J.~C.~V.\ \& Pringle, J.~E. 1985, MNRAS, 217,
387

\reference{}Papaloizou, J.~C.~B.\ \& Spruit, H.~C. 1993, ApJ, 417, 677.

\reference{}Patterson, J., Jablonsky, F., Koen, C., O'Donoghue, D.\ \&
Skillman, D.~R. 1995, PASP, 107, 1183

\reference{}Pojmanski, G. 1986, Acta Astron., 36, 69

\reference{}Pringle, J.~E. 1981, ARA\&A, 19, 137

\reference{}Rozyczka, M. \& Spruit, H.~C. 1993, ApJ, 417, 677

\reference{}Semeniuk, I. 1980, A\&AS, 39, 29

\reference{}Shakura, N.~I.\ \& Sunyaev, R.~A. 1973, A\&A, 24, 337

\reference{}Smak, J. 1984, Acta Astr., 34, 93

\reference{}Smak, J. 1987, A\&SS, 131, 497

\reference{}Smak, J. 1996, in Cataclysmic Variables and Related Objects,
eds.\ A.~Evans, \& J.~H.\ Wood (Dordrecht: Kluwer), p.~45

\reference{}Tout, C.~A.\ \& Pringle, J.~E. 1992, MNRAS, 259, 604

\reference{}Vishniac, E.~T. 1997, ApJ, 482, 414

\reference{}Vishniac, E.~T.\ \& Wheeler, J.~C. 1996, ApJ, 471, 921

\reference{}Vogt, N. 1974, A\&A, 36, 369

\reference{}Warner, B. 1987, MNRAS, 227, 23

\reference{}Warner, B. 1995, Cataclysmic Variable Stars (Cambridge:
Cambridge University Press)

\reference{}Whitehurst, R. 1988, MNRAS, 232, 35

\reference{}Whitehurst, R.\ \& King, A. 1991, MNRAS, 249, 25

\reference{}Wood, J., Horne, K., Berriman, G., Wade, R., O'Donoghue, D.,
\& Warner, B. 1986, MNRAS, 219, 629

\reference{}Wood, J.~H. et~al. 1989, MNRAS, 239, 809

\end{references}
\end{document}